\documentclass[aps,prb,reprint,superscriptaddress,noeprint]{revtex4-2}
\usepackage{siunitx}
\usepackage{graphicx}			

\usepackage{color}
\usepackage{orcidlink}

\begin{document}
\newcommand{\avec}{\textbf{\textit{a}}}
\newcommand{\bvec}{\textbf{\textit{b}}}
\newcommand{\cvec}{\textbf{\textit{c}}}
\newcommand{\Hvec}{\textbf{\textit{H}}}
\newcommand{\krc}{K$_2$ReCl$_6$}
\newcommand{\ksc}{K$_2$SnCl$_6$}
\newcommand{\CsCrCl}{Cs$_3$Cr$_2$Cl$_9$}
\newcommand{\rucl}{RuCl$_3$}
\newcommand{\nio}{Na$_2$IrO$_3$}


\newcommand{\tfrac}[2]{{\footnotesize \frac{#1}{#2}}} 
\newcommand{\boldsymbol}[1]{\textbf{\textit{#1}}} 

\title{Interplay of magnetic order and ferroelasticity  in the spin-orbit coupled antiferromagnet K$_2$ReCl$_6$}

\author{Alexandre Bertin\,\orcidlink{0000-0001-5789-3178}}
\email{bertin@ph2.uni-koeln.de}
\affiliation{II. Physikalisches Institut, Universit\"at zu K\"oln, Z\"ulpicher Str. 77, 50937 K\"oln, Germany}
\author{Tusharkanti Dey\,\orcidlink{0000-0002-3244-0452}}
\altaffiliation{now at: Department of Physics, IIT (ISM) Dhanbad, Jharkhand 826004, India}
\affiliation{II. Physikalisches Institut, Universit\"at zu K\"oln, Z\"ulpicher Str. 77, 50937 K\"oln, Germany}
\author{Daniel Br\"uning\,\orcidlink{0000-0001-5830-2683}}
\affiliation{II. Physikalisches Institut, Universit\"at zu K\"oln, Z\"ulpicher Str. 77, 50937 K\"oln, Germany}
\author{Dmitry Gorkov\,\orcidlink{0000-0002-5983-2771}}
\affiliation{II. Physikalisches Institut, Universit\"at zu K\"oln, Z\"ulpicher Str. 77, 50937 K\"oln, Germany}
\affiliation{Heinz Maier-Leibnitz Zentrum (MLZ), Technische Universit\"at M\"unchen, Lichtenbergstr. 1, 85748 Garching, Germany}
\author{Kevin Jenni}
\affiliation{II. Physikalisches Institut, Universit\"at zu K\"oln, Z\"ulpicher Str. 77, 50937 K\"oln, Germany}
\author{Astin Krause}
\affiliation{II. Physikalisches Institut, Universit\"at zu K\"oln, Z\"ulpicher Str. 77, 50937 K\"oln, Germany}
\author{Petra Becker\,\orcidlink{0000-0003-4784-3729}}
\affiliation{Abteilung Kristallographie, Institut f\"ur Geologie und Mineralogie, Universit\"at zu K\"oln, Z\"ulpicher Str. 49b, 50674 K\"oln, Germany}
\author{Ladislav Bohat\'y}
\affiliation{Abteilung Kristallographie, Institut f\"ur Geologie und Mineralogie, Universit\"at zu K\"oln, Z\"ulpicher Str. 49b, 50674 K\"oln, Germany}
\author{Daniel Khomskii}
\affiliation{II. Physikalisches Institut, Universit\"at zu K\"oln, Z\"ulpicher Str. 77, 50937 K\"oln, Germany}
\author{Vladimir Pomjakushin\,\orcidlink{0000-0003-2180-8730}}
\affiliation{Laboratory for Neutron Scattering and Imaging, PSI, CH-5232 Villigen PSI, Switzerland}
\author{Lukas Keller\,\orcidlink{0000-0002-8492-4117}}
\affiliation{Laboratory for Neutron Scattering and Imaging, PSI, CH-5232 Villigen PSI, Switzerland}
\author{Markus Braden\,\orcidlink{0000-0002-9284-6585}}
\email{braden@ph2.uni-koeln.de}
\affiliation{II. Physikalisches Institut, Universit\"at zu K\"oln, Z\"ulpicher Str. 77, 50937 K\"oln, Germany}
\author{Thomas Lorenz\,\orcidlink{0000-0003-4832-5157}}
\email{tl@ph2.uni-koeln.de}
\affiliation{II. Physikalisches Institut, Universit\"at zu K\"oln, Z\"ulpicher Str. 77, 50937 K\"oln, Germany}

\date{\today}

\begin{abstract}

The magnetic and structural phase transitions occurring in K$_2$ReCl$_6$ were studied by macroscopic and microscopic techniques.
Structural phase transitions associated with rotations of the ReCl$_6$ octahedra lower the symmetry from cubic to monoclinic,
form ferroelastic domains, and are visible in susceptibility, specific heat and thermal expansion measurements.
In the antiferromagnetically ordered state slightly below $T_{\rm N}$=12\,K these domains can be rearranged by a magnetic field
inducing a relative elongation of the polydomain crystal parallel to the field of 0.6\%.
At zero field the magnetic structure in K$_2$ReCl$_6$ does not exhibit a weak ferromagnetic component, but at large magnetic field
a distinct magnetic structure with a finite weak ferromagnetic component is stabilized.
High magnetic fields rearrange the domains in the crystal to align the weak ferromagnetic moment parallel to the field.
The altered domain structure with the crystal elongation is abruptly suppressed at lower temperature but persists upon heating to
well above $T_{\rm N}$. However, heating above the lowest structural phase transition and successive cooling restore the initial
shape.

\end{abstract}

\pacs{}

\maketitle

\section{Introduction}

Materials with strong spin-orbit coupling (SOC) currently attract strong interest in various fields of condensed matter research, because it can give rise to fascinating new states of matter with highly anomalous properties. For example, topological insulators often arise from materials with strong SOC~\cite{TI_RevModPhys.83.1057}, as well as magnetic skyrmions~\cite{Fert2017}, and SOC is the source of the anomalous Hall effect, spin Hall effect, or Rashba coupling~\cite{AHE_RevModPhys.82.1539,Spin-Hall_PhysRevLett.83.1834,Rashba_JETP1984}. Concerning the search of quantum spin liquids, one major focus is on materials which may realize the Kitaev model~\cite{Kitaev2006}. 
{ This exactly solvable model was formulated for a spin-1/2 honeycomb lattice with bond-directional Ising-type exchange interaction, 
and again, SOC is the key which may yield an (approximate) realization of this model in some materials when a total spin $S=1/2$ couples with an effective orbital moment $\ell_{eff}=1$ to an effective $j_{eff}=\ell_{eff}-S=1/2$.
Meanwhile, many extensions of the original model were subject of detailed theoretical works and revealed that Kitaev-type spin liquid states can be realized in various regions of an extended parameter space commonly labeled Heisenberg-Kitaev-$\Gamma$ model. This model also includes $j_{eff}=1/2$ states on an fcc lattice, which can be found in double-perowskites, e.g.\ Ba$_2$CeIrO$_6$~\cite{Aczel2019, Revelli2019}, or in halides of the antifluorite structure type, e.g.\ K$_2$IrCl$_6$~\cite{Khan2019,Reig-i-Plessis2020}.}

In a wider context, it is of interest to study the general impact of a strong SOC on  the magnetic, electronic and structural properties of heavy transition-metal compounds. In fact, experimental evidence for multipole order has been reported for some $A_2$TaCl$_6$ materials for different alkaline-earth ions $A^+$ and Ta$^{4+}$ with $t_{2g}^{1}$ configuration~\cite{Ishikawa2019}. 
Moreover, a recent theoretical study has shown that the strength of SOC determines whether a partially filled $d$ shell is Jahn-Teller active or not. In particular a $d$ shell with three electrons in the $t_{2g}$ orbitals that is non-degenerate for weak SOC can become Jahn-Teller distorted in the $jj$-coupling scheme~\cite{Streltsov2020}.

Motivated by these considerations we studied K$_2$ReCl$_6$ with the Re$^{4+}$ ions in such a $t_{2g}^{3}$ electronic configuration. At room temperature this material is cubic, but shows a sequence of structural phase transitions upon cooling and finally antiferromagnetic (AFM) order evolves at $T_{\rm N} \simeq 12\,$K~\cite{Busey1962,Busey1962a} Although \krc\ has been subject of numerous studies in the past~\cite{Busey1962,Busey1962a,OLeary1970, Armstrong1980,Willemsen1977,Willemsen1977a}, the crystal structure of the low-temperature phases was not yet determined. This also holds for the magnetic structure, because the existing neutron diffraction studies~\cite{Smith1966,Minkiewicz1968} do not take into account the reduced symmetry of the low-temperature phases, but refer to the room-temperature face-centered cubic structure. 
	
Here, we report thermodynamics and diffraction experiments on \krc , which reveal outstandingly strong magnetoelastic anomalies. In the antiferromagnetically ordered phase, the application of a magnetic field of about 10\,T causes a sample elongation by up to 0.6\% that can be traced back to a magnetic-field-induced reorientation of ferroelastic structural domains. 
The domain reorientation persists when the sample is heated well above $T_{\rm N}$, but the initial shape is recovered when the sample is heated above a structural phase transition at $77\,$K. The field-induced domain reorientation correlates with the occurrence of a weak ferromagnetic moment. Combining this macroscopic information with zero-field neutron diffraction experiment and symmetry analysis allows us to conclude that the magnetic space group of \krc\ changes as a function of the magnetic field. 

The paper is organized as follows: After introducing the experimental techniques in Sec.~\ref{exp}, we present the macroscopic studies across the three structural and the magnetic phase transitions in Sec.~\ref{mac}. Then, we discuss the crystal structure data of \krc\ with particular emphasis on the controversial issues about the sequence of structural phase transitions in previous reports~\cite{OLeary1970,Lynn1978,Noda1980,Kugler1983,Ihringer1984}. In addition, the relations between the directions 
in macroscopic measurements on partially (de-)twinned samples and the lattice parameters of the different low-temperature tetragonal and monoclinic phases are clarified. In Sec.~\ref{MS}, we discuss the  magnetization and expansion data showing huge magnetoelastic anomalies at large magnetic fields. These anomalies can be traced to the reorientation of magnetostructural domains that is related to a field-dependent change of the magnetic structure as it is discussed in Sec.~\ref{mag}.

\section{Experimental details}
\label{exp}

Transparent green high-quality single crystals of \krc\ were grown from hydrochloric acid solution by controlled, slow evaporation of the solvent at 295\,K.
These crystals were cut to samples with typical dimensions in the mm range with oriented [1\,0\,0]$_c$ or [1\,1\,1]$_c$ directions of the room-temperature cubic phase.
The subscript '$c$' will be used throughout this paper whenever the notation refers to the cubic lattice of the high-temperature phase with lattice parameter $a\approx 9.8\,$\AA.
Magnetization and heat capacity were measured using commercial magnetometer (SQUID and VSM) and relaxation-time calorimeter setups, respectively (MPMS and PPMS, Quantum Design). The uniaxial thermal expansion along the [1\,0\,0]$_c$ direction was measured with a capacitance dilatometer~\cite{kuchler2012} mounted in the variable-temperature insert of a 14-Tesla cryomagnet (Oxford Instruments).

A crushed sample was measured in Bragg-Brentano geometry on a D5000 diffractometer using Cu $K_\alpha$ radiation. The sample was cooled in a He-evaporation cryostat and a small amount of Si was added to calibrate the instrument. The diffraction patterns were analyzed by the Rietveld method using the \textsc{Fullprof} program package~\cite{Carvajal1993}. 
A comparison between experimental and calculated diffraction patterns is given in the Supplemental Material~\cite{supplmat}. 

Single-crystal X-ray diffraction experiments were conducted on a Bruker AXS Kappa Apex II four-circle diffractometer with Mo$K_{\alpha}$ radiation ($\lambda=0.71$~\AA), in combination with an Oxford N-Helix cryosystem for the low temperatures.

The temperature dependence of magnetic Bragg reflections were studied on the KOMPASS spectrometer at the Forschungsreaktor Munich II in Garching~\cite{Kompass}.
A monochromatic beam with a wavelength of 4~\AA \ was generated with a highly oriented pyrolytic graphite monochromator and higher orders were
suppressed with a velocity selector. 
The crystal was mounted in the cubic (1\,0\,0)$_c$/(0\,1\,1)$_c$ scattering geometry in a cryostat using a closed-cycle refrigerator.
In order to determine the zero-field magnetic structure we performed powder neutron-diffraction experiments on the DMC diffractometer at the SINQ neutron source of the Paul Scherrer Institute. 5.7\,g of a commercial sample powder from Alfa Aesar filled a 8\,mm diameter\,$\times$\,50\,mm height cylindrical Vanadium sample holder and mounts in an ILL-orange cryostat. The wavelength was set to 2.45\,\AA .
The instrument resolution file reflecting the exact geometry of the experimental setup was calibrated with a powder sample of Na$_2$Ca$_3$Al$_2$F$_{14}$ and used in the refinements with the  \textsc{Fullprof}  program package~\cite{Carvajal1993}.

\section{Results and discussion}

\subsection{Macroscopic measurements across the three structural and magnetic transitions}
\label{mac}

\begin{figure}[t]
	\includegraphics[width=0.96\columnwidth]{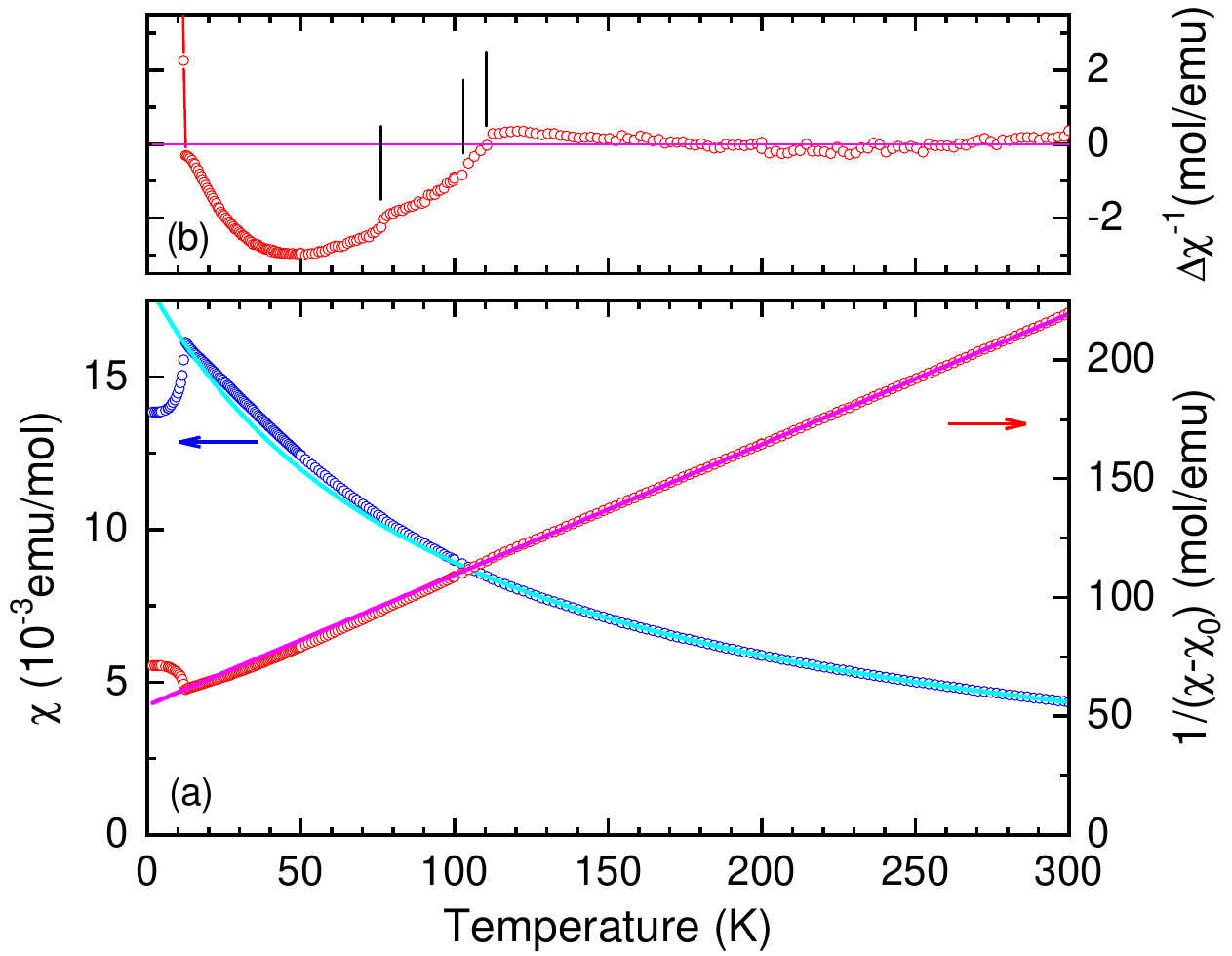}
	\caption{(a) Magnetic susceptibility $\chi $ (blue circles) of \krc\ measured in a magnetic field $\mu_0 H=0.1\,$T applied along the [1\,1\,1]$_c$ direction of the high-temperature cubic phase. The kink at $T_{\rm N} \simeq 12\,$K signals AFM order.		
		Red data points are the inverse susceptibility after subtracting $\chi_0 \simeq -2\times 10^{-4}$emu/mol due to the diamagnetic contributions of inner shells~\cite{Bain2008}. The red and blue lines show a Curie-Weiss curve obtained by fitting the data for $T>140\,$K. (b) The difference between fit and $1/(\chi - \chi_0)$ data reveals distinct anomalies at each of the three structural phase transitions (see Fig.~\ref{fig:cp}(a)) marked by the vertical bars.}
	\label{fig:chi}
\end{figure}

\begin{figure}[t]
	\includegraphics[width=0.96\columnwidth]{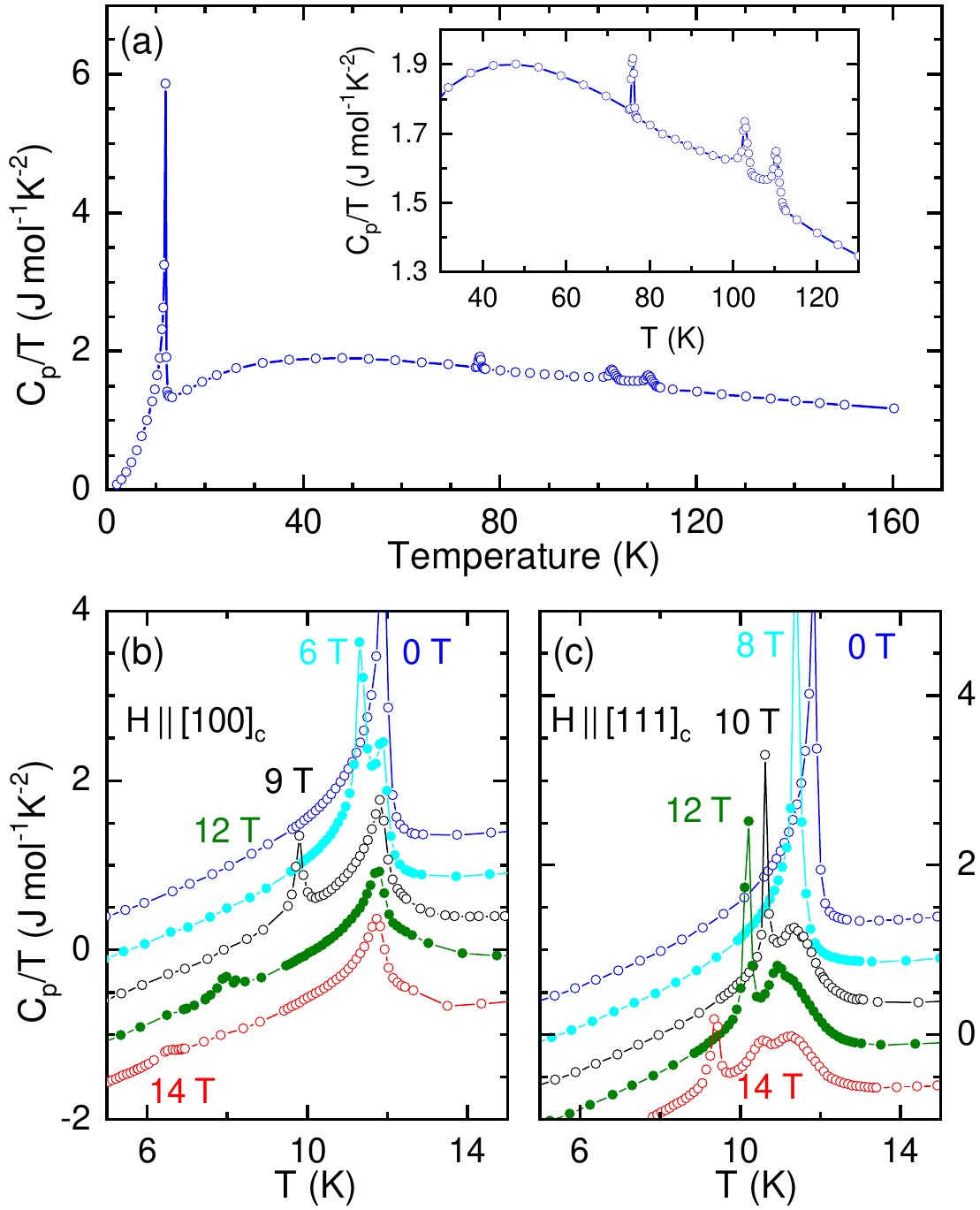}
	\caption{(a) ZF specific heat of \krc\ showing distinct peaks in $C_{p}/T$, which signal AFM order at $T_{\rm N} \simeq 12\,$K and three structural transitions at $\simeq 77\,$K, $103\,$K, and $111\,$K~\cite{Busey1962,Busey1962a,OLeary1970}. The latter are displayed in an expanded view in the inset. (b) and (c) $C_{p}/T$ measured in magnetic fields applied along the cubic [1\,0\,0]$_c$ and [1\,1\,1]$_c$ axes, respectively. For clarity, the curves for different fields are shifted by $-0.5\,$J$\,$mol$^{-1}$K$^{-2}$ with respect to each other.}
	\label{fig:cp}
\end{figure}

Figure~\ref{fig:chi}(a) displays the magnetic susceptibility $\chi $ obtained in a magnetic field $\mu_0 H=0.1\,$T applied along the [1\,1\,1]$_c$ direction of the \krc .
The abrupt, but weak, decrease below  $T_{\rm N} \simeq 12\,$K signals AFM order with magnetic moments possessing only a small component parallel to the magnetic field direction.
This agrees with neutron results of the zero-field (ZF) magnetic structure revealing that the ordered moments are aligned within cubic (1\,0\,0)$_c$  planes~\cite{Smith1966,Minkiewicz1968}. 
It is important, however, that upon cooling \krc\ undergoes a sequence of three structural phase transitions at $\simeq 111\,$K, $103\,$K, and $77\,$K, see Fig.~\ref{fig:cp}(a)~\cite{Busey1962,Busey1962a,OLeary1970, Armstrong1980}, but the microscopic character and the symmetry of the distortions remain unknown.
The low-temperature structure is monoclinic, see below, but may appear as pseudo-cubic due to different structural domains, as will be further discussed below. In the paramagnetic phase, the inverse susceptibility is well described by a Curie-Weiss behavior:
\begin{equation}\label{eq:CW}
\chi(T)= N_{\rm A} \frac{\mu_{eff}^2}{3k_{\rm B}(T-\theta)}+\chi_0 \, ,
\end{equation}

where, $N_{\rm A}$ ($k_{\rm B}$) is Avogadro's (Boltzmann's) constant, $\mu_{eff}$ the effective magnetic moment, $\theta$ the Weiss temperature and $\chi_0 \simeq -2\times 10^{-4}$emu/mol is a temperature independent background estimated from the core diamagnetism of K$^{+}$, Re$^{4+}$, and Cl$^{-}$~\cite{Bain2008}. Because of the structural phase transitions, the fit is restricted to $T>140\,$K. The difference $\Delta \chi^{-1}$, see Fig.\ref{fig:chi}(b),  between fit and data reveals small but distinct anomalies at each of the structural phase transitions. The fit yields a Weiss temperature $\theta \simeq -99\,$K, i.e.\ an overall AFM coupling, and $\mu_{eff}\simeq 3.816\,\mu_{\rm B}$ which agrees within $2\,$\% to the expected value for local moments of $3\,\mu_{\rm B}$\footnote{A free fit yields $\chi_{0}\simeq -4.2\times 10^{-4}$emu/mol, $\theta\simeq-112\,$K, $\mu_{eff}\simeq 3.98\,\mu_{\rm B}$, but such a strongly negative $\chi_0$ appears unreasonable. Previous results for \krc\ are surprisingly rare and yield $\mu_{eff}\approx 3.7\,\mu_{\rm B}$ and $\theta\approx -60\,$K~\cite{Nelson1954,Busey1962}.}. This would suggest spin-only magnetism of the Re$^{4+}$ ions with $S=3/2$ and the large frustration ratio $|\theta|/T_{\rm N}\simeq 8$ is related to their fcc arrangement with nearest-neighbor (NN) AFM coupling. Note, however, that the structural distortions cause an enhancement of $\chi$ and
thus a reduction of the total AFM interaction, see Fig. 1. These results roughly agree with earlier analyses reporting $\theta=-55(5)\,$K and $\mu_{eff}\simeq 3.63\,\mu_{\rm B}$~\cite{Busey1962}.

The ZF specific-heat data shown in Fig.~\ref{fig:cp}(a) reveal pronounced anomalies at each phase transition in agreement with previous work~\cite{Busey1962a}. The three structural phase transitions are associated with rotations of the octahedra, and such transitions are quite common for this structure type~\cite{OLeary1970,Boysen1978,Lynn1978,Armstrong1980,Reig-i-Plessis2020,Khan2021,Stein2023}.
The influence of large magnetic fields on $T_{\rm N}$ is shown in panels (b) and (c). For $H || [1\,0\,0]_c$, we observe a splitting into an upper $T_{\rm N1}$ with a very weak field dependence ($\simeq -0.01\,$K/T) and a lower $T_{\rm N2}$, which is strongly field-dependent and the corresponding anomaly is significantly broadened above $9\,$T. For $H || [1\,1\,1]_c$, there is again a splitting, which requires larger fields and is more complex because the corresponding anomalies are of different shapes and the highest field data even indicate three transitions. In order to interpret these data, one has to consider the occurrence of different domains at the magnetic and at the various structural transitions, as will be discussed in sections~\ref{Xray} and~\ref{mag}.

\begin{figure}[t]
	\includegraphics[width=0.98\columnwidth]{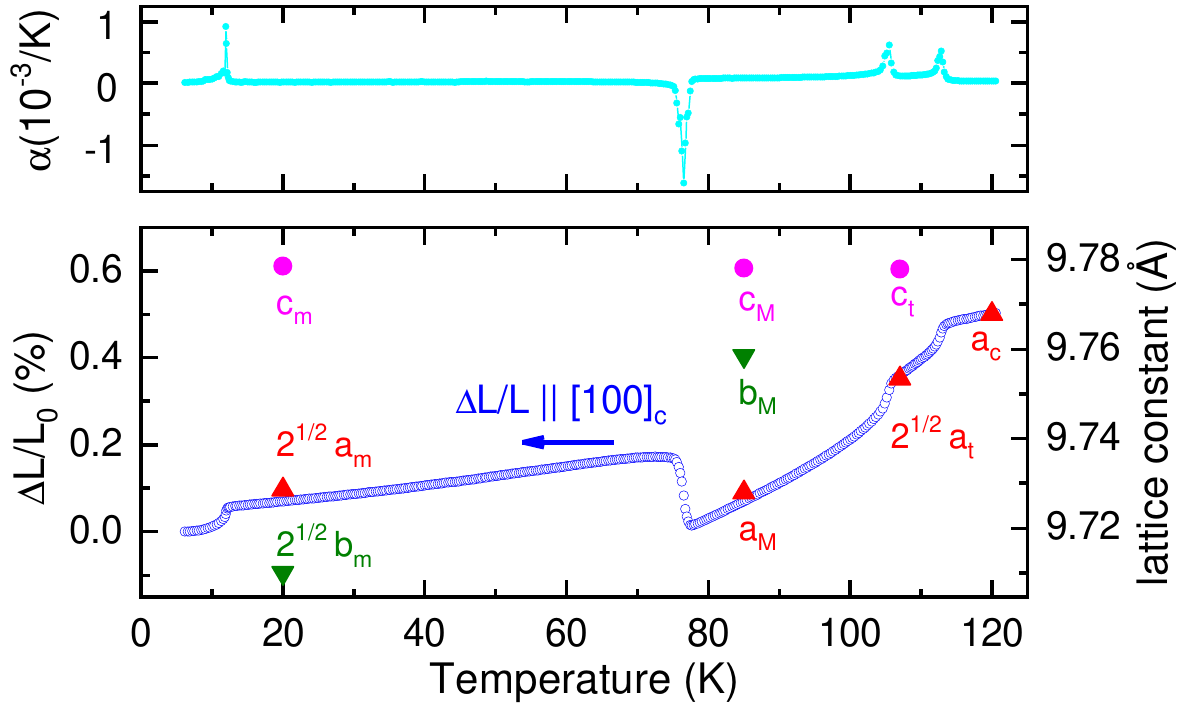}
	\caption{Uniaxial thermal expansion of \krc\ measured along one of the $[1\,0\,0]_c$ directions of the high-temperature cubic phase. The lower panel shows the relative length change $\Delta L(T)/L_0$ in dark blue and the corresponding uniaxial thermal expansion coefficient $\alpha$ is displayed in the upper panel. For comparison, the lower panel also contains the lattice constants obtained from X-ray diffraction data in the various low-temperature phases of \krc , see Table~\ref{tab:character_table}.}
	\label{fig:dlzero}
\end{figure}

Figure~\ref{fig:dlzero} shows the uniaxial thermal expansion along a $[1\,0\,0]_c$ direction, which reveals pronounced anomalies at each of the transitions. The strongest anomaly is found at the $77\,$K transition, which causes a quasi-discontinuous increase of $\Delta L/L_0$ identifying this transition to be of first order, in agreement with previous studies~\cite{OLeary1970, Willemsen1977,Armstrong1980}. The structural transitions at 103~K and 111\,K and also the magnetic transition at $12\,$K  are of second order and cause continuous spontaneous contractions upon cooling. The asymmetric shape of the respective anomalies of the thermal expansion coefficient $\alpha=1/L_0 \, \partial \Delta L/\partial T $ signals strong fluctuations, which were analyzed for analogous ZF expansion data measured along the cubic $[1\,1\,1]_c$ direction~\cite{Willemsen1977,Willemsen1977a}. 

Figure\;\ref{fig:dlzero} also compares $\Delta L(T)/L_0$ with the temperature dependent lattice constants of the various low-temperature phases that will be discussed in detail in the following Sec.\;\ref{Xray}. Here, we mainly concentrate on the cubic-to-tetragonal transition at 111\,K and use the information from the X-ray data that this transition causes a spontaneous expansion of the tetragonal $c_t$ axis whereas the tetragonal in-plane axes $a_t$ shrink and are rotated by 45$^\circ$ with respect to the  cubic $a_c$ axes. Consequently, the cubic $[1\,0\,0]_c$ direction, along which the macroscopic length change $\Delta L(T)/L_0$ is measured, can either represent the thermal expansion of the tetragonal axis $c_t$, or of one of the tetragonal in-plane directions $[\pm 1\,1\,0]_t$. In general, such type of ferroelastic transitions tend to form differently orientated twin domains depending on which of the three cubic $a_c$ axes becomes the tetragonal axis $c_t$. The multidomain nature of \krc\ is known from previous studies~\cite{OLeary1970,Armstrong1980} and can be clearly seen by optical microscopy on thin single crystals as is shown in the supplemental material~\cite{supplmat}. Thus, one could expect that the macroscopic $\Delta L(T)/L_0$ is given by a superposition of tetragonal in-plane and out-of plane directions, but the comparison in Fig.\;\ref{fig:dlzero} shows that $\Delta L(T)/L_0$ very well matches the relative difference between $a_c$ and $\sqrt{2}\, a_t$ measured at 120\,K and 104\,K, respectively (see Table\;\ref{tab:character_table}). This means that the macroscopic $\Delta L(T)/L_0$ is measured on a partially detwinned sample, which mainly consists of domains with $c_t$ axes pointing along one of the two possible orientations perpendicular to the measured cubic $[1\,0\,0]_c$ direction. This can be explained by a finite uniaxial pressure along the measurement direction, because in the used capacitance dilatometer the sample is clamped by CuBe springs with a force of about $3\,$N~\cite{kuchler2012} that results in a uniaxial pressure of about $0.5\,$MPa on the studied sample. Such pressures usually do not affect the measured length changes of solids, unless there are phase transitions with soft (magneto)structural domains~\cite{Niesen2013,Niesen2014a,Kunkemoller2016,Kunkemoller2017,Bruening2021}. In \krc , cooling under $0.5\,$MPa uniaxial pressure along one of the $a_c$ axes is sufficient to efficiently suppress the occurrence of tetragonal domains with their long $c_t$ axis along this direction. Upon further cooling, analogous domain orientations with their shortest directions along the uniaxial pressure direction are obtained in the different low-temperature monoclinic phases as will be discussed in the following section. In Sec.\,\ref{MS} we will show that magnetic fields cause relative length changes of up to 0.6\,\% that signal a magnetic-field induced domain switching because this $\Delta L/L_0$ agrees well with the relative difference between the lattice constant $c_{m}$ and the sum $(a_{m}+b_{m})/\sqrt{2}$ of the low-temperature monoclinic phase.

\subsection{Powder X-ray diffraction studies across the structural phase transitions}
\label{Xray}

The high-temperature crystal structure of \krc \ corresponds to the cubic space group $Fm\bar{3}m$ with a lattice constant of 9.8\AA~\cite{Smith1966}.
Thermal expansion measurements along [1\,1\,1]$_c$ \cite{Willemsen1977,Willemsen1977a} and [1\,1\,0]$_c$ and temperature dependent X-ray diffraction studies of the (8\,0\,0) reflection indicate the reduction of the
symmetry to a tetragonal lattice at the 111\,K transition, and to at least an orthorhombic lattice at the 77\,K transition~\cite{OLeary1970,Armstrong1980}.
Also NMR and NQR studies indicate this loss of symmetry~\cite{OLeary1970,Armstrong1980}.
In addition, inelastic neutron scattering experiments find a softening of rotational phonon modes
at the center, $\Gamma$, and at the $X$ point of the Brillouin zone~\cite{Lynn1978}.
This neutron experiment further suggests that the 111\,K transition is associated
with the zone-center mode and thus does not result in a breaking of translation symmetry.
This conclusion agrees with the temperature dependence of the (2\,1\,3)$_c$ X-ray superstructure
reflection violating the $F$ centering only below the 77\,K transition \cite{OLeary1970}, but our
data clearly exclude this scenario.
In spite of these various studies revealing the occurrence of structural phase transitions ~\cite{Busey1962,Busey1962a,OLeary1970, Armstrong1980}, 
the exact crystal structure of this material has not been determined for its low-temperature phases. Most likely, the twinning in the tetragonal and monoclinic
phases prevented a quantitative analysis of single-crystal data at these times.

\begin{figure}[b]
\includegraphics[width=.96\columnwidth]{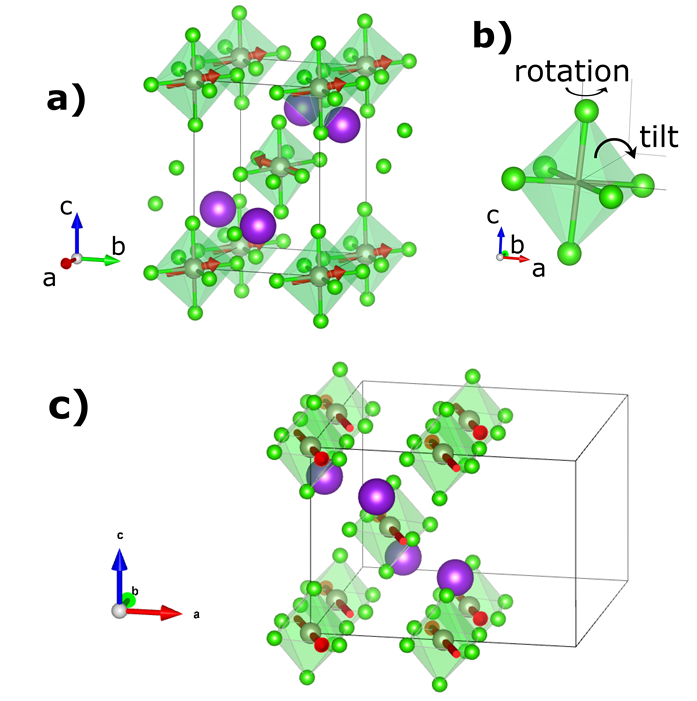}
\caption{(a) Crystal structure of \krc\ in the low-temperature monoclinic phase (labeled by $m$) with space group $P2_1/n$. Dark green, light green, and purple spheres represent Cl, Re, and K ions, respectively. (b) An isolated ReCl$_6$ octahedron with the relevant rotation axes to illustrate the used nomenclature. 
In $P2_1/n$, the octahedra rotate around the $c_m$ axis with a phase shift between octahedra at (0,0,0) and (0.5,0.5,0.5) while the tilt of the octahedra around roughly $b_m$ is in-phase at all sites. Note that $c_m$ corresponds to the cubic lattice constant, while $a_m$ and $b_m$ are reduced by a factor close to $\sqrt{2}$ due to the rotation of axes. The arrows in panel (a) illustrate the high-field magnetic structure described by the propagation vector (0,0,1) which exhibits an AFM component along $b_m$ and a canting of moments resulting in weak ferromagnetism along $c_m$. The canting has been exaggerated for better visibility. (c) Zero-field magnetic structure with the propagation vector (0.5,0.5,0) causing a doubling of magnetic lattice as indicated by thin lines. Pictures were drawn with the visualization software~\textsc{VESTA3}~\cite{Momma2011}.}
	\label{fig:struc}
\end{figure}

The crystal structure of \krc\ is shown in Fig.~\ref{fig:struc} in its low-temperature modification with monoclinic symmetry.
The so-called K$_2$PtCl$_6$ structure type is characterized by octahedra that are
not interconnected in contrast to the perovskites with corner-sharing octahedra.
Due to this greater mutual independence, one may expect structural instabilities due to rotation of the octahedron that were analyzed
with respect to the size of the $A$ ion by Brown \cite{Brown1964}.
For K$_2$SeBr$_6$ Noda {\it et al.}~\cite{Noda1980} reported the following sequence of structural phases~\footnote{Authors of Ref.~\cite{Noda1980} refer to the $C2/c$ space group with the non-standard setting $C2/n$}:
\begin{equation}\label{eq:spgr}
 Fm\bar{3}m \rightarrow P4/mnc \rightarrow C2/c \rightarrow P2_1/n
\end{equation}

These successive transitions can be explained by soft phonon modes~\cite{Lynn1978,Armstrong1980}.
The condensation of the rotation around a single axis defined by a Re-Cl bond, see Fig.~\ref{fig:struc}(b), 
results in a tetragonal space group: either $P4/mnc$ if the $X$ mode condenses, or $I4/m$ if the $\Gamma$ mode condenses, and each of these transitions into a tetragonal phase can be continuous. However, breaking of translation symmetry occurs in $P4/mnc$ and thus superstructure reflections should appear, while the primitive
cell in $I4/m$ is identical to that of the high-temperature $F$-centered phase.
For K$_2$SeBr$_6$~\cite{Noda1980} and for K$_2$SnCl$_6$~\cite{Boysen1978,Kugler1983,Ihringer1984}, the $X$ mode
is reported to condense first, and thus breaking translation symmetry, while the neutron scattering and early X-ray diffraction studies on \krc\
concluded the opposite~\cite{Lynn1978,OLeary1970}.

\begin{table*}
	\caption{\label{tab:character_table} Symmetries of the different structural phases and corresponding lattice parameters determined by powder X-ray diffraction. The lower part indicates the equivalences of the cubic main symmetry directions in the distorted phases and the indexation of the (100)$_c$ reflection in the low
		symmetry lattices. For the directions we also indicate the multiplicities in the various space groups.}
	\begin{ruledtabular}
		\begin{tabular}{c|c|c|c}
			$Fm\bar{3}m$         &    $P4/mnc$                     &    $C2/c$    & $P2_1/n$ \\
			120\,K & 104\,K & 85\,K & 20\,K \\ \hline
			$a$=9.7677(3)~\AA     &       $a_{t}=6.8968(2)$~\AA                         &     $a_{M}=9.7277(4)$~\AA           &  $a_{m}=6.8800(5)$~\AA  \\
			$b$=$a$              &       $b_{t}$=$a_t$                              &     $b_{M}=9.7594(6)$~\AA           &  $b_{m}=6.8657(6)$~\AA  \\
			$c$=$a$              &       $c_{t}=9.7784(6)$~\AA                         &     $c_{M}=9.7779(5)$~\AA           &  $c_{m}=9.7788(5)$~\AA \\
			$\beta$=90$^{\circ}$              &        $\beta$=90$^{\circ}$                                   &      $\beta_{M}$=90.05(1)$^{\circ}$           &    $\beta_{m}$=90.08(1)$^{\circ}$   \\ 
			{[100]$_c$\,6$\times$}  & [110]$_t$\,4$\times$, [001]$_t$\,2$\times$ & [100]$_M$\,2$\times$, [010]$_M$\,2$\times$, [001]$_M$\,2$\times$  &   [110]$_m$=[$\bar{1}$10]$_m$\,4$\times$, [001]$_m$\,2$\times$  \\
			{[110]$_c$\,12$\times$ }& [100]$_t$\,4$\times$, [112]$_t$\,8$\times$ & [110]$_M$\,4$\times$, [101]$_M$\,2$\times$, [10$\bar{1}$]$_M$\,2$\times$, [011]$_M$\,4$\times$ & [200]$_m$\,2$\times$, [020]$_m$\,2$\times$, [111]$_m$\,4$\times$, [11$\bar{1}$]$_m$\,4$\times$\\
			{[111]$_c$\,8$\times$}  & [101]$_t$\,8$\times$  & [111]$_M$\,4$\times$, [11$\bar{1}$]$_M$\,4$\times$  & [201]$_m$\,2$\times$ [20$\bar{1}$]$_m$\,2$\times$, [021]$_m$\,4$\times$\\
			(100)$_c$         &   (110)$_t$ (001)$_t$ &           (100)$_M$ (010)$_M$ (001)$_M$         & (001)$_m$ (0.5 -0.5 0)$_m$$\equiv$(0.5 0.5 0)$_m$ \\ 
		\end{tabular}
	\end{ruledtabular}
\end{table*}

\begin{figure}[b]
	\includegraphics[width=\columnwidth]{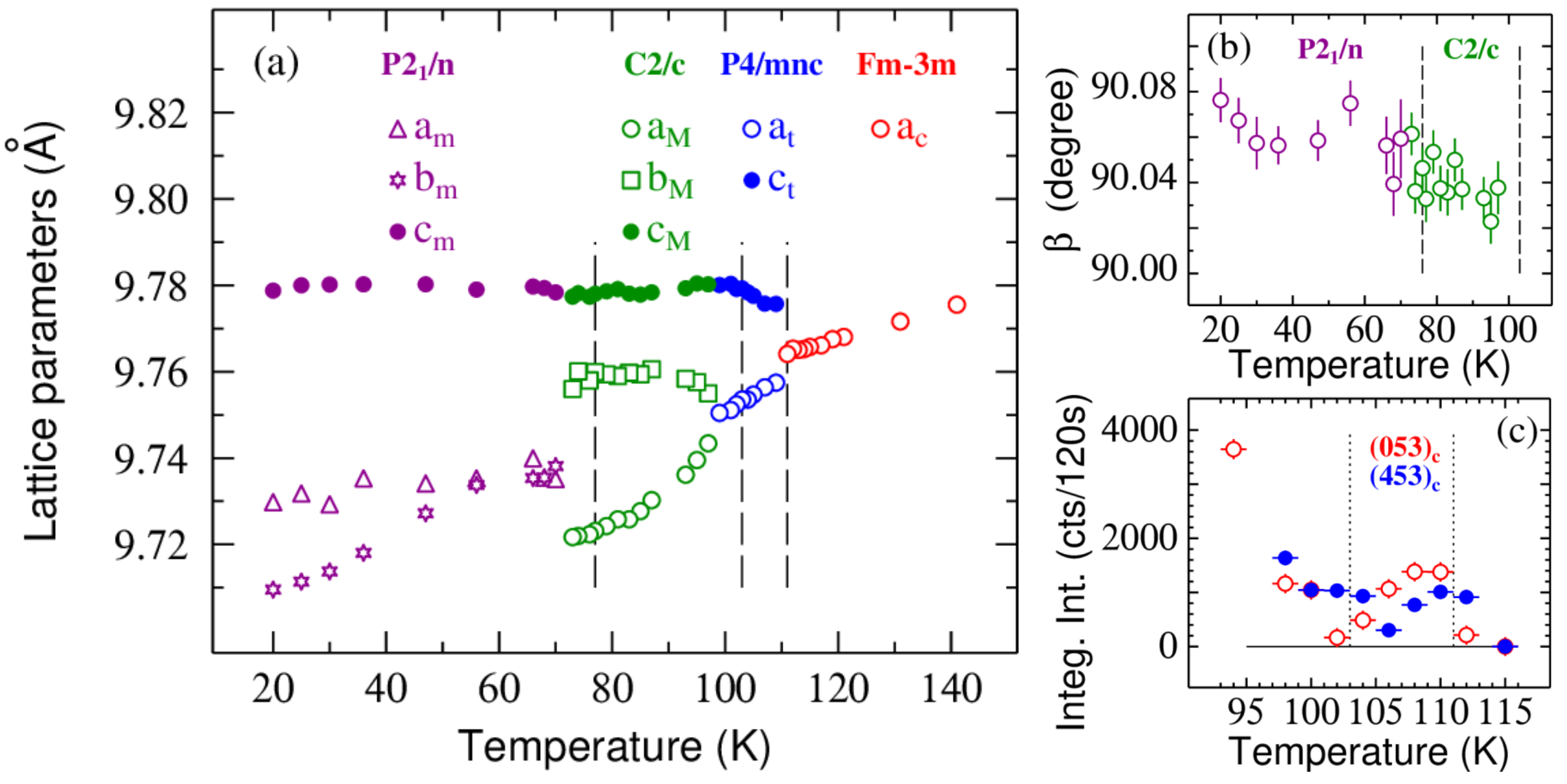}
	\caption{(a) Temperature dependent lattice parameters in \krc\ across the structural phase transitions indicated by dashed lines. Different colors are used for the 
		different structural phases and the lattice parameters $a_t$ of $P4/mnc$ and $a_m, b_m$ of $P2_1/n$ are multiplied by $\sqrt{2}$ to account for the rotation of these axes. (b) Evolution of the monoclinic angle $\beta$ in the $C2/c$ and $P2_1/n$ phases. (c) Integrated intensity of the superstructure reflections (3\,5\,0)$_c$ and (4\,5\,3)$_c$ across the transitions into the $P4/mnc$ and $C2/c$ phases. Here, the 115~K data of the cubic phase have been subtracted. Because of the gas flow cooling method of the Oxford N-Helix cryosystem, a sample temperature uncertainty of $\pm$1~K is estimated.}
	\label{fig:latt}
\end{figure}

We have therefore studied (3\,5\,0)$_c$ and (4\,5\,3)$_c$ superstructure reflections as a function of temperature by single-crystal X-ray diffraction, see Fig.~\ref{fig:latt}(c).
These reflections clearly appear already at the highest transition at $\simeq 111$\,K, which thus can be associated with the translation symmetry breaking $X$ mode,
yielding $P4/mnc$ symmetry. Our analysis disagrees with the earlier study of the (2\,1\,3)$_c$ reflection~\cite{OLeary1970}. Note, however, that the
multiple domains can easily explain the heating-induced loss of superstructure intensity at a lower temperature, while our observation
of the persistence of translation symmetry breaking intensities up to 111\,K is unambiguous. The inelastic neutron scattering results
on the phonons~\cite{Lynn1978} were interpreted as evidence for the $\Gamma$ mode to condense first,
but within their error bars the data are fully consistent with the opposite sequence, as given in Eq.~(\ref{eq:spgr}).

All our single-crystal diffraction studies on crystals with dimensions of about 100\,$\mu$m reveal the occurrence of structural twinning, as it is expected for a continuous ferroelastic structural phase transition. The emergence of twins mimics point-group symmetries that are lost at the structural phase transitions. For \krc , the twin domains can be visualized by polarized light microscopy and form regular patterns~\cite{supplmat}, as it is typically observed for such  ferroelastic domains~\cite{Sapriel_PhysRevB.12.5128,Khachaturyan1991_PRB}.

The combination of two rotation modes around the bond axes but with
different stacking (one corresponding to an $X$ and the other to a $\Gamma$ mode) results in the monoclinic
space group $C2/c$ and the transition into this structure at $\simeq 103$\,K can be continuous for each of the two tetragonal phases discussed above.
In order to distinguish these distortions, we denote the $X$ distortion as rotation and the in-phase $\Gamma$ scheme as tilt [see Fig.~\ref{fig:struc}(b)].
The transition from $C2/c$ to $P2_1/n$ corresponds to a change of the tilt axis. In $C2/c$ the tilt also occurs around
a Re-Cl bond, while in $P2_1/n$ the tilt occurs essentially around an axis parallel to an octahedron edge as it is illustrated in Fig.~\ref{fig:struc}(b).
The combination of such rotation and tilt structural instabilities is also characteristic for three-dimensional~\cite{Carvajal1998,Cwik2003,zhou2005} or layered
perovskites~\cite{braden1998}, and a temperature-dependent change of a tilt scheme is also well documented, e.g.\ in some cuprates~\cite{Axe}.
The phase transition between $C2/c$ and $P2_1/n$ must be of first order since the tilt axis changes abruptly. Note that the first-order character of the corresponding 77\,K transition of \krc\ reported from 
previous studies~\cite{OLeary1970, Willemsen1977,Armstrong1980} is also reflected by the quasi-discontinuous $\Delta L(T)/L_0$ of our thermal-expansion data, see Fig.~\ref{fig:dlzero}, which in addition 
shows a pronounced temperature hysteresis, as will be seen in Sec.~\ref{MS}.

We have studied the crystal structure of \krc\ by temperature dependent powder X-ray diffraction. 
Although these experiments cannot resolve the positional parameters in \krc\ with sufficient precision to discuss the possible splitting of the Re-Cl bond distances, the lattice parameters including the tetragonal and monoclinic splittings are precisely determined and given in Fig.~\ref{fig:latt}(a).
The temperature dependence of the structural parameters obtained with the powder X-ray diffraction is given in the supplemental material~\cite{supplmat}.
Powder X-ray diffraction cannot distinguish the two possible
tetragonal space groups due to the weakness of the superstructure reflections, but the tetragonal splitting appearing below 111\,K is unambiguous. The lattice elongates along the rotation axis which is
the tetragonal axis $c_t$, whereas the tetragonal in-plane axes $a_t$ shrink yielding an overall weak reduction of the lattice volume in agreement with thermal expansion along [1\,1\,1]$_c$~\cite{Willemsen1977,Willemsen1977a}.
The measurements clearly determine the two distinct monoclinic symmetries, although the monoclinic angle deviates very little from 90$^{\circ}$.
In $C2/c$, the monoclinic lattice constants (labeled by a  subscript '$M$') correspond to those of the high-temperature cubic phase. Compared to the tetragonal phase, one recognizes a strong splitting of $a_M$ and $b_M$ that is expected due to the onset of tilting around a bond axis (close to $b_M$).
This proves that the monoclinic phase occurring at 103\,K indeed corresponds to a rotation and a tilt of the octahedra around the Re-Cl bonds.
In contrast, the low-temperature monoclinic phase in space group $P2_1/n$ (labeled by the  subscript '$m$') exhibits a reduced splitting of the monoclinic axes $a_m$ and $b_m$, which are rotated by 45$^{\circ}$ with respect to the original cubic axes. The monoclinic lattice below 77\,K is clearly incompatible with a tilt around a Re-Cl bond
but requires the tilt around an axis more parallel to the octahedron edge as illustrated in Fig.~\ref{fig:struc}(b). The temperature dependence of the lattice parameters indicates that the tetragonal splitting emerging at the highest phase transition is not qualitatively changing at the lower transitions, which is difficult to reconcile with the initial suggestion stating that the upper transition would correspond to the $\Gamma$ mode distortion and would then change at the lowest transition. Thus, the powder and the single-crystal X-ray diffraction studies confirm that the general sequence of phase transitions given in Eq.~\ref{eq:spgr} applies to \krc. The main symmetry properties and the relations of directions in the various structural phases are summarized in Table~\ref{tab:character_table}.

The temperature dependent lattice constants from Fig.\;\ref{fig:latt}(a) have been partially included already in Fig.\;\ref{fig:dlzero} for the above discussion of the macroscopic expansion data $\Delta L(T)/L_0$ measured along a cubic $[1\,0\,0]_c$ direction. Upon cooling from 120 to 85\,K, the measured $\Delta L(T)/L_0$ first follows the temperature-dependent evolution from the cubic $a_c(T)$ axis to the tetragonal in-plane direction [1\,1\,0]$_t$ and then to the axes $a_M(T)$ of the upper monoclinic phase. These directions correspond to the respective shortest directions of the distorted phases that are parallel to a $[1\,0\,0]_c$ direction, see Table\;\ref{tab:character_table}. At the 77\,K transition into the low-temperature monoclinic phase, the measured $[1\,0\,0]_c$ length jumps to the length of the monoclinic diagonal, $(a_{m}+b_{m})/\sqrt{2}$, which again represents the shortest distance of the monoclinic lattice that can align parallel to $[1\,0\,0]_c$. As discussed above, this reveals that the uniaxial pressure of about 0.5\,MPa to clamp the sample in the used dilatometer is sufficient to induce domain arrangements for the tetragonal and monoclinic phases, where the shortest lattice directions are aligned parallel to the pressure direction.	
In the low-temperature monoclinic phase this is prohibited by the fact that the two shorter directions $a_m$ and $b_m$ are rotated by 45$^{\circ}$ relative to the cubic axes, and, consequently, the domains align such that the longest $c_m$ axes is perpendicular to the pressure direction.

The upper two structural phase transitions are ferroelastic, see supplemental material~\cite{supplmat}, and this allows to control the corresponding domains through strain for both transitions. Note that the $ Fm\bar{3}m \rightarrow P4/mnc $ transition is an improper ferroelastic transition because the translation symmetry is also broken, whereas the  $P4/mnc \rightarrow C2/c$ is a proper ferroelastic transition~\cite{stokes1988}. In contrast, the $C2/c \rightarrow P2_1/n$ transition at 77\,K is not ferroic, but the structural change accompanying the onset of antiferromagnetic order in zero magnetic field is again ferroelastic, see Sec.~\ref{mag}.

\subsection{Strong magnetoelastic anomaly and weak ferromagnetism}
\label{MS}

Despite the occurrence of 12 monoclinic twin domain orientations that were observed in the single-crystal X-ray diffraction experiments,
the cubic $[1\,0\,0]_c$ axis corresponds to superpositions of only two symmetry-inequivalent directions of the low-temperature monoclinic phase, namely
${\bf c}_{m}$ and $({\bf a}_{m}\pm {\bf b}_{m})$. Note that for a given $[1\,0\,0]_c$ direction there are four monoclinic domain orientations possible which have ${\bf c}_{m} \| [1\,0\,0]_c$ and there are eight domains with $({\bf a}_{m}\pm {\bf b}_{m}) \| [1\,0\,0]_c$.
For the following discussion of the thermal expansion and of the magnetization data, which both were measured in magnetic fields up to 14~T applied along a cubic $[1\,0\,0]_c$ axis, we can ignore that these domains are additionally slightly rotated against the cubic axes and against each other.

\begin{figure}[t]
	\includegraphics[width=0.98\columnwidth]{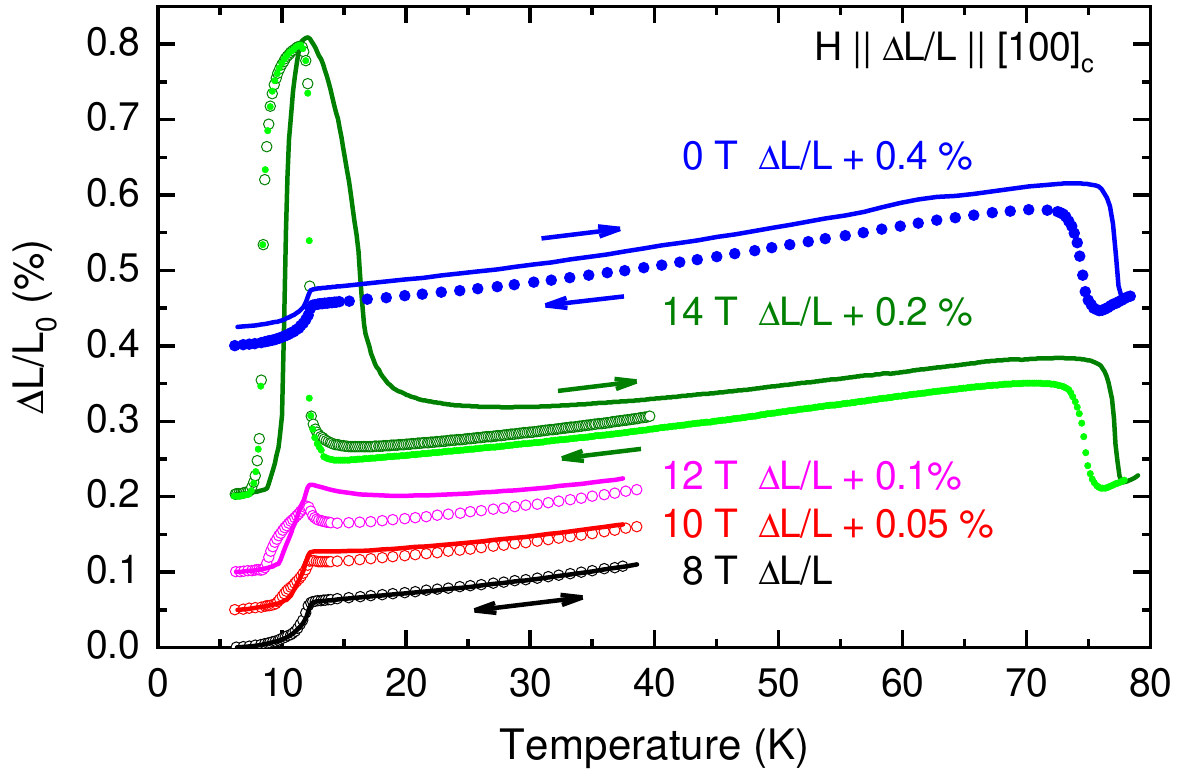}
	\caption{A series of uniaxial thermal expansion measurements along the high-temperature cubic [100]$_c$ direction for different longitudinal fields $H \parallel \Delta L$. The field was stepwise increased from 8\,T to $14\,$T at $40\,$K, and $\Delta L(T)/L_0$ was then measured with decreasing ($\circ$) and increasing (---) temperature. In $14\,$T, the sample was heated above the structural transition and field-cooled ($\bullet$) again to $6\,$K. Then the field was switched off and an analogous ZF temperature cycle of $\Delta L(T)/L_0$ was measured. For clarity, the curves for different fields are shifted with respect to each other as marked in the figure.}
	\label{fig:dl}
\end{figure}

Figure~\ref{fig:dl} displays $\Delta L(T)/L_0$ curves obtained for different magnetic fields. Starting with a field of 8\,T applied at 40\,K, $\Delta L(T)/L_0$ was measured upon cooling to $6\,$K and heating back to $40\,$K. The field was increased in steps of $2\,$T, and analogous thermal cycles were repeated. After the 14~T measurement, the field was switched off at 6~K and a zero-field thermal cycle followed. In $8\,$T, $\Delta L(T)/L_0$ is fully reversible and hardly differs from the ZF data, whereas a systematic difference between $\Delta L(T)/L_0$ measured upon cooling and heating evolves in $10\,$T and becomes more obvious in $12\,$T. As was already observed in the heat capacity data, see Fig.\ref{fig:cp}, the transition splits into an upper $T_{\rm N1}$ that remains close to the ZF $T_{\rm N}\simeq 12\,$K and a lower $T_{\rm N2}\approx 9\,$K. The lower $T_{\rm N2}$ is hysteretic and, moreover, the thermal cycle back to $40\,$K results in a plastic elongation of $\Delta L/L_0 \simeq 0.02\,$\%. These features become drastically enhanced during the thermal cycle in $14\,$T, where the intermediate phase around $11\,$K exceeds the neighboring low- and high-temperature phases even by $\Delta L/L_0 > 0.5\,$\%. With increasing temperature this plastic deformation continuously decreases to $\Delta L/L_0 \simeq 0.04\,$\% above about $40\,$K and vanishes completely at the first-order structural transition around  $77\,$K. The subsequent field-cooling process in $14\,$T results in a slightly reduced  $\Delta L/L_0$ (light-green symbols) compared to the previous field-cooling run (dark-green symbols), but both curves merge below $T_{\rm N1}\simeq 12\,$K suggesting that the same intermediate- and low-temperature phases are reached below $T_{\rm N1}$ and $T_{\rm N2}$, respectively. Reducing the field back to zero at $6\,$K results in a ZF phase with a plastic elongation $\Delta L/L_0 \simeq 0.03\,$\%, which again vanishes at the $77\,$K transition as is seen during the subsequent ZF thermal cycle of $\Delta L(T)/L_0$.

The drastic elongation of $\Delta L/L_0  > 0.5\,$\% around 11\,K and 14\,T strongly exceeds typical magnetostrictive effects arising from exchange striction. Instead, it perfectly matches the microscopically determined relative difference between the lattice constant $c_{m}$ and the sum $(a_{m}+b_{m})/\sqrt{2}$ of the low-temperature monoclinic phase revealing that we are dealing with an essentially complete switching from $({\bf a}_{m}\pm {\bf b}_{m})$-oriented structural domains to an almost completely ${\bf c}_{m}$-oriented crystal.
This switching is magnetic-field induced in the magnetically ordered state, but it does not necessarily change upon heating above $T_{\rm N}$, because the structural domains are not directly sensitive to the magnetic field in the paramagnetic phase.
Due to their ferroelastic nature, the structural domains are affected by the uniaxial pressure in the dilatometer, and one can expect a temperature-activated
switching back to the shorter $({\bf a}_{m}\pm {\bf b}_{m})$-oriented domains. This naturally explains the hysteresis above $T_{\rm N}$, because the obtained structural domain distribution depends on the maximum temperature and remains frozen in the subsequent cooling process. After heating above the first-order structural phase transition at 77\,K, an analogous domain distribution is obtained when the sample is cooled down to the ordered phase under analogous external conditions. In the present case, this is essentially a pure $({\bf a}_{m}\pm {\bf b}_{m})$-oriented state which remains unchanged down to $T< T_{\rm N}$ where a spontaneous (staggered) magnetization evolves, at least for fields below about 8\,T. However, cooling in larger magnetic fields, e.g.\ 10 to 14\,T, induces a more or less complete switching to the elongated ${\bf c}_{m}$-oriented domains at the upper $T_{\rm N1}$, followed by another domain reversal back to $({\bf a}_{m}\pm {\bf b}_{m})$-oriented domains at $T_{\rm N2}$.   

\begin{figure}[t]
	\includegraphics[width=0.98\columnwidth]{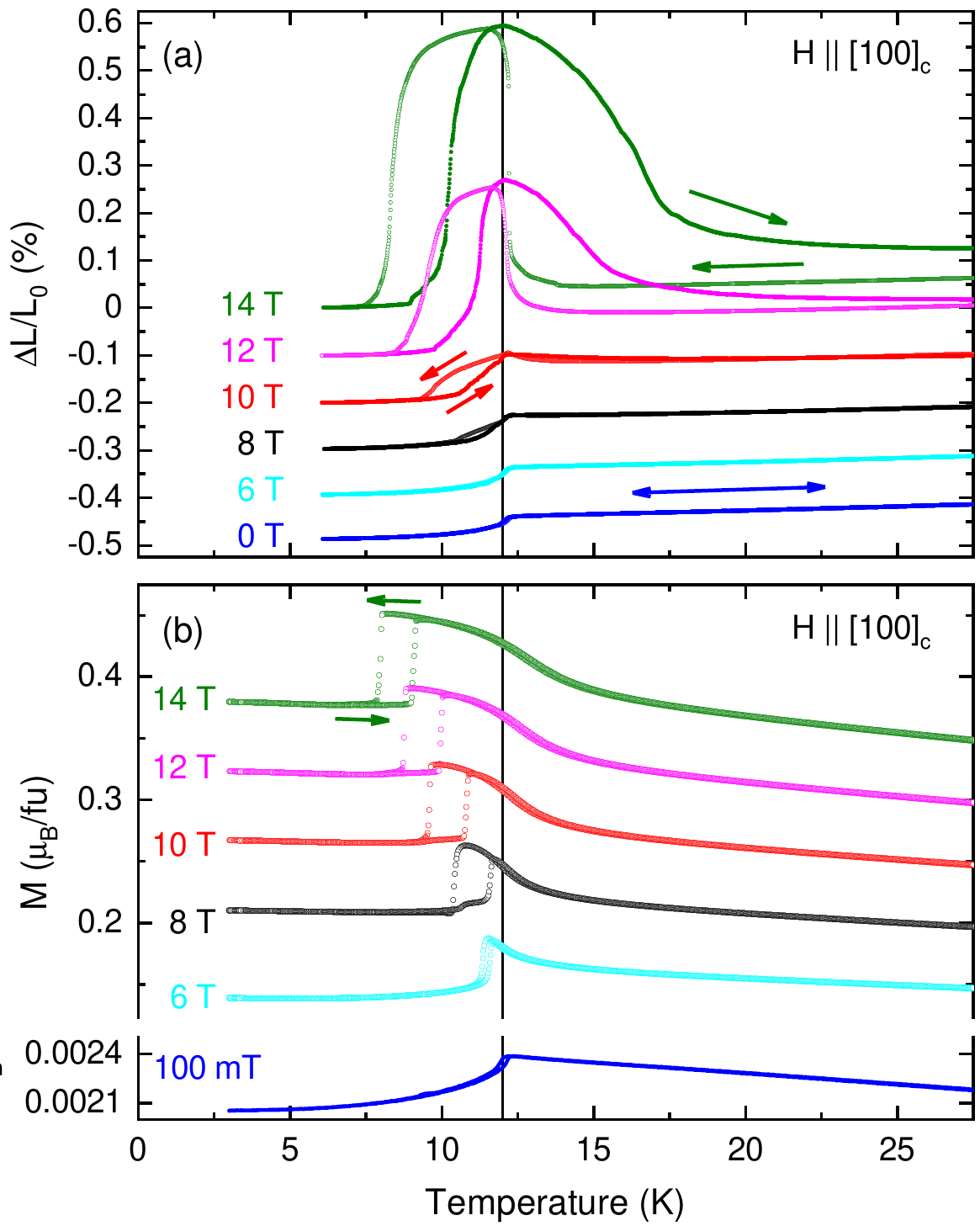}
	\caption{(a) Uniaxial expansion along the [100]$_c$ direction of the high-temperature cubic phase for various fields $H \parallel \Delta L/L_0 \parallel [100]_c$. The field was step-wise decreased from $14\,$T at $40\,$K, and for each field $\Delta L(T)/L_0$ was measured upon cooling to 6~K and heating to 40\,K, as indicated by arrows. The curves for different fields are offset by $-0.1\,$\% with respect to each other for clarity. (b) Analogous temperature cycles between $30$~K and $3\,$K of the magnetization $M(T)$ where the field was again step-wise decreased from $14\,$T at 30~K. Note the different scale for the 100\,mT data. The vertical line marks the ZF $T_{\rm N}\simeq 12\,$K.}
	\label{fig:dl_mag}
\end{figure}

In Fig.~\ref{fig:dl_mag}(a) we show another set of thermal expansion measurements in comparison to a corresponding set of magnetization data $M(T)$ displayed in Fig.~\ref{fig:dl_mag}(b). Here, we started with the highest field of 14\,T applied in the paramagnetic state and measured $\Delta L(T)/L_0$ or $M(T)$ upon cooling to, respectively, 6\,K or 3\,K, and upon heating back to about 35\,K. Then analogous temperature cycles were repeated after decreasing the field in steps of typically 2\,T. For the highest field of 14\,T, the $\Delta L(T)/L_0$ curves of Fig.~\ref{fig:dl_mag}(a) agree well with the corresponding data of Fig.~\ref{fig:dl}, but in the intermediate field range of 8 to 12\,T we now observe  significantly larger hystereses between the $\Delta L(T)/L_0$ curves measured with decreasing or increasing temperature. The enhanced hysteresis for a given field signals a memory effect, because the structural domain distribution in the intermediate field range depends on details of the previous temperature cycle, in particular, whether the previous cycle has been performed in a stronger or in a smaller magnetic field.

\begin{figure}[t]
	\includegraphics[width=0.98\columnwidth]{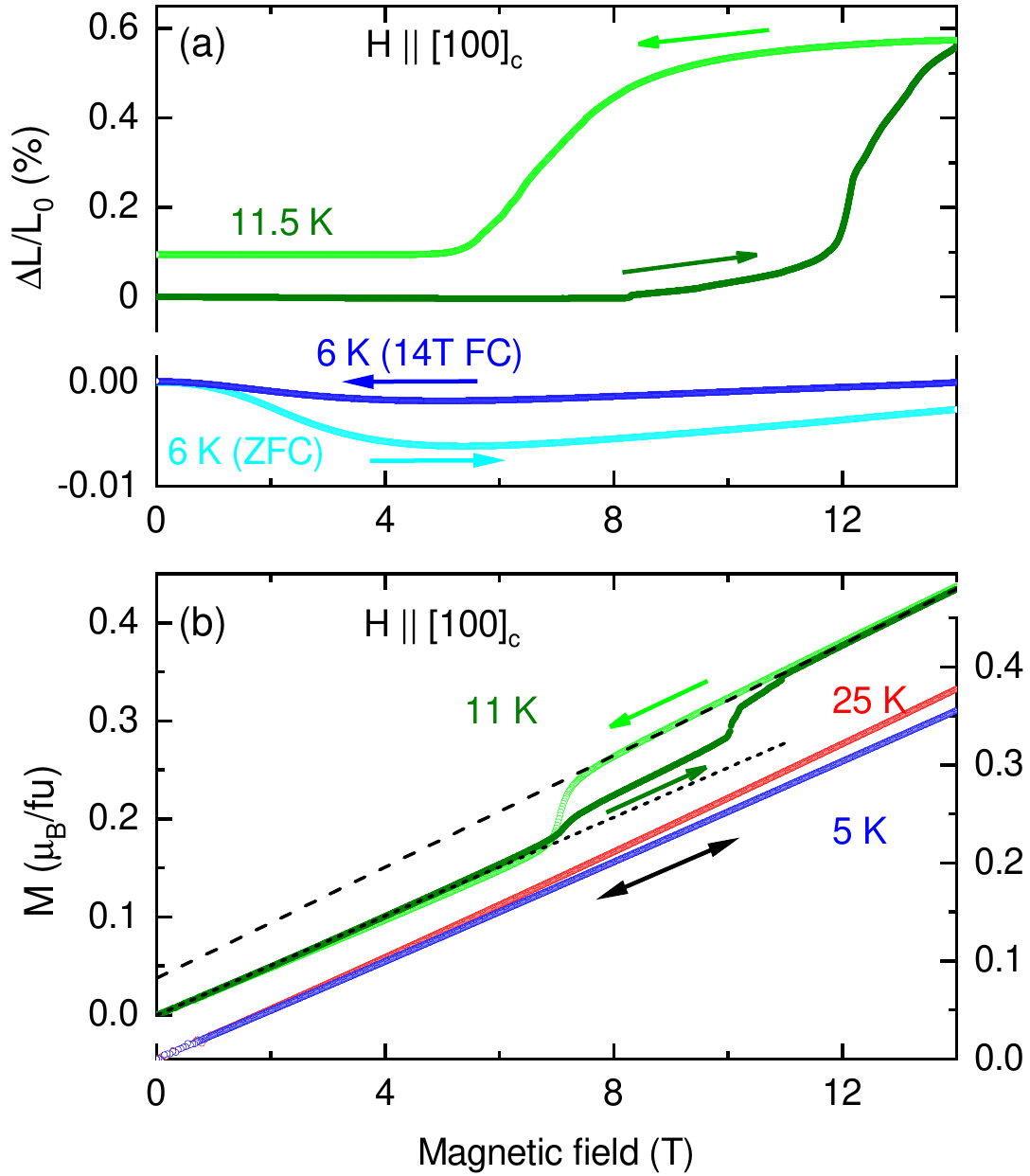}
	\caption{(a) Uniaxial expansion as a function of a longitudinal magnetic field measured along the [100]$_c$ direction of the high-temperature cubic phase, i.e. $\Delta L(H)/L_0 || H|| [100]_c$. The hysteresis loop at 11.5\,K signals domain reorientation, which is not fully reversible. For comparison, two separate  $\Delta L(H)/L_0$ curves are shown, which both are measured at 6\,K, either after cooling in zero field or in a field of 14\,T. Note the different $\Delta L(H)/L_0$ scale.  (b) Magnetization curves $M(H)$ for different temperatures measured with increasing and decreasing field $H || [100]_c$. The dotted (dashed) line is a linear fit of the low-(high-)field range of the field-in(de)creasing run of the 11\,K curve.}
	\label{fig:dl_mag_vs_H}
\end{figure}

The most prominent feature of the corresponding $M(T)$ cycles, see Fig.~\ref{fig:dl_mag}(b), are first-order transitions with sharp magnetization jumps and a pronounced hysteresis. From 14\,T down to 8\,T, the position and width ($\approx$ 1\,K) of the $M(T)$ hysteresis fairly agree with the corresponding hysteresis in $\Delta L(T)/L_0$. The quantitative differences could result from the fact that the two data sets do not stem from exactly the same sample. However, the strong hysteresis seen in $\Delta L(T)/L_0$ above $T_{\rm N}$ is absent in $M(T)$. Moreover, the 6\,T curves of both data sets show qualitatively different transitions. These differences can be attributed to the competing influences of a magnetic field versus uniaxial pressure on the (re)orientation of structural domains in combination with a weak ferromagnetism resulting from canted staggered moments, as will be explained below. 

Figure~\ref{fig:dl_mag_vs_H} displays magnetic-field dependent data of $\Delta L(H)/L_0$ and $M(H)$, which both reveal a first-order transition around 11\,K with a broad hysteresis region from about 7 to 12\,T. Panel~(a) also
shows that the elongation at this transition exceeds the field-induced  $\Delta L(H)/L_0$ within the low-temperature phase by two orders of magnitude. As discussed, these drastic length changes signal the switching between ${\bf c}_{m}$- and $({\bf a}_{m}\pm {\bf b}_{m})$-oriented structural domains. From the non-reversible part of $\Delta L(H)/L_0$ we can estimate a domain switching ratio of up to about 80\,\%. As shown in Fig.~\ref{fig:dl_mag_vs_H}(b), the $M(H)$ curves at 5 and 25\,K are strictly linear with very similar slopes. This is typical for a transverse-field configuration of an antiferromagnet: the ordered staggered moments at 5\,K are almost perpendicular to the magnetic field and continuously cant towards the field direction with increasing $H$. At 11\,K, this remains unchanged in the low-field phase, whereas in the high-field phase, $M(H)$ is still linear but now with a finite ZF offset of $\simeq 0.04\,\mu_{\rm B}/$fu, which signals weak ferromagnetism resulting from slightly canted staggered moments. The ratio of this offset to the local moment of $3\,\mu_{\rm B}/$fu derived from the Curie-Weiss fit yields a canting angle of about $0.8^\circ$. The weak ferromagnetism is consistent with the high-field $M(T)$ curves shown in Fig.~\ref{fig:dl_mag}(b), which upon cooling from the paramagnetic phase continuously increase until the first-order transition is reached and the magnetization shows a jump-like decrease of about $0.04-0.07\,\mu_{\rm B}/$fu depending on the magnetic field. This magnetization stays well below the value of a ferromagnetic state with
local moments of several $\mu_{\rm B}$. Therefore, this signal must stem from a weak ferromagnetic component that is coupled to the
staggered antiferromagnetic order of much larger magnetic moments. The magnetization data reveal that the  first-order transitions are related to the appearance and disappearance of weak ferromagnetism whereas the expansion data show that this transition is related to the switching of structural domains. This means that the weak ferromagnetic moment is pointing mainly parallel to the long $c_{m}$ axis and, vice versa, the staggered magnetic moments are lying almost within the monoclinic $ab$ planes with a small out-of-plane canting of the order of $1^\circ$.

For  $H \parallel [100]_c$, the weak ferromagnetism favors to align the $c_{m}$ axis along $[100]_c$, but this competes with the uniaxial pressure in the capacitance dilatometer which favors to align the $({\bf a}_{m}\pm {\bf b}_{m})$ domains along $[100]_c$. Thus, the $\Delta L(T)/L_0$ curves mainly reflect the behavior of the  $({\bf a}_{m}\pm {\bf b}_{m})$ axis for $H \parallel ({\bf a}_{m}\pm {\bf b}_{m})$. Up to 10~T, this holds for the entire temperature range, while above 12\,T it only holds below $\sim$7\,K, and in the paramagnetic phase only for $\Delta L(T)/L_0$ obtained upon cooling. All those $\Delta L(T)/L_0$ curves show anomalies close to the ZF $T_{\rm N}\simeq 12\,$K, which are naturally interpreted as indicating the onset of the staggered spontaneous magnetization. This onset is almost field-independent, but above 10\,T it is combined with an abrupt domain reorientation from $({\bf a}_{m}\pm {\bf b}_{m}) \parallel H$ to ${\bf c}_{m} \parallel H$.

For the magnetization measurements a $mm$-sized rectangular sample was glued with one side face to a flat sample holder such that there is hardly any pressure on the sample. In low field, one can thus expect a rather homogeneous distribution of structural domains depending on microscopic details such as internal strains, whereas  a ${\bf c}_{m}$-oriented domain arrangement will be induced by a large $H \parallel [100]_c$ around 11\,K. Due to the absence of opposing uniaxial pressure, the ${\bf c}_{m}$-oriented domains can be reached more easily and remain more stable upon heating into the paramagnetic state. Thus, the high-field $M(T)$ curves in Fig.~\ref{fig:dl_mag_vs_H}(b) represent the $c_{m}$-axis magnetization, which continuously evolves upon crossing the transition from the paramagnetic phase to the canted AFM phase with a weak ferromagnetic moment. At a lower temperature, however, the weak ferromagnetic moment disappears via a first-order transition. In the magnetization data, this transition is systematically sharper than in the corresponding $\Delta L(T,H)/L_0$ curves. This correlates with the absence or presence of a sizable uniaxial pressure in the respective experimental setup, indicating that the bare disappearance of the weak ferromagnetism at low temperature leaves the ${\bf c}_{m}$-oriented domain state essentially unchanged, in analogy to heating up into the paramagnetic phase.

The weak ferromagnetic phase exists only in a small pocket in the $B$--$T$ phase diagram of \krc . As can be inferred from Figs.~\ref{fig:dl}, \ref{fig:dl_mag}, and \ref{fig:dl_mag_vs_H}, it appears above about 8\,T close to the zero-field $T_{\rm N}$ and upon further cooling it disappears again via a hysteretic first-order transition. This explains the observation of two transition temperatures in the specific heat data, see Fig.~\ref{fig:cp}, and with further increasing field the width of this pocket shows a weak, but systematic increase.

\subsection{Magnetic structure}
\label{mag}

According to early neutron scattering data~\cite{Smith1966,Minkiewicz1968}, the magnetic structure of \krc\ consists of ferromagnetic cubic (1\,0\,0)$_c$ planes with the magnetic moments of adjacent planes being aligned antiparallel to each other, and ordered moments of $2.6(5)\,\mu_{\rm B}$ and $2.7(3)\,\mu_{\rm B}$ per Re$^{3+}$ ion were derived in Refs.~\cite{Smith1966} and~\cite{Minkiewicz1968}, respectively. 
However, a complete determination of the magnetic structure taking 
into account the reduced symmetry of the real low-temperature structure of \krc\ has not been published.

\begin{figure}[t]
	\centering
	\includegraphics[width=\columnwidth]{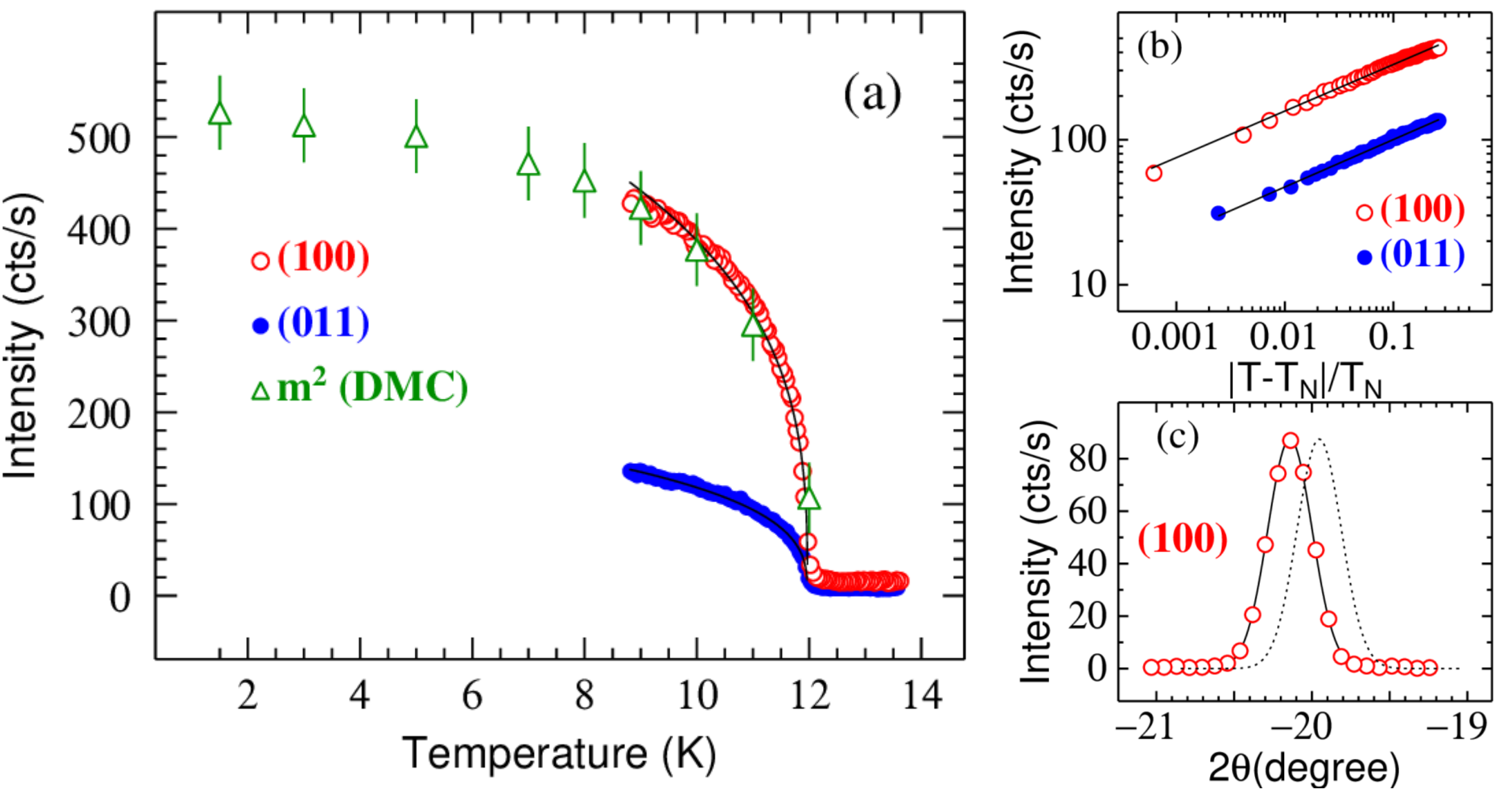}
	\caption{Temperature dependent integrated intensities of the magnetic (1\,0\,0)$_c$ and (0\,1\,1)$_c$ reflections on linear (a) and on double logarithmic (b) scales. The temperature dependence of $m^2$ deduced from the powder neutron diffraction experiment on DMC is also added, with these data scaled to the (1\,0\,0)$_c$ data point measured at 9K. (c) The profile of a scan across the magnetic (1\,0\,0)$_c$ reflection versus 2$\theta$. The data were fitted by a Gaussian distribution (solid line) and the comparison to a simulation based on the averaged lattice constant (dotted line) unambiguously reveals that the magnetic propagation vector corresponds to the shortened distance in the monoclinic lattice.
	}
	\label{magn-peaks}
\end{figure}

Figure~\ref{magn-peaks}(a) shows the intensities of the (1\,0\,0)$_c$ and (0\,1\,1)$_c$ magnetic Bragg reflections as a function of temperature, which we measured at the cold neutron triple-axis spectrometer KOMPASS at the MLZ. The temperature dependence can be well described by a simple power law with respect to the reduced temperature: $I=I_0\left(\frac{T_{\rm N}-T}{T_{\rm N}}\right)^{\beta}$ yielding $T_{\rm N}$=11.97(2)~K and $\beta$=0.321(8) for (1\,0\,0)$_c$ and $T_{\rm N}$=11.97(2)~K and $\beta$=0.327(6) for (0\,1\,1)$_c$, see Fig.~\ref{magn-peaks}(b). The critical exponent agrees very well with a three dimensional Ising model~\cite{Pelissetto02}. By adding 10 minutes collimators before and after the sample, the position of the magnetic (1\,0\,0)$_c$ peak is analyzed as a function of the scattering angle, see Fig.~\ref{magn-peaks}(c), and clearly reveals that it corresponds to the shorter distance of the monoclinic lattice, because it appears at a larger absolute 2$\theta$ value compared to the expected position from the average lattice constant.

The propagation vector (1\,0\,0)$_c$ in the cubic lattice, which is observed in our and in earlier neutron diffraction, can be either a (0\,0\,1)$_m$ or a (0.5\,0.5\,0)$_m$ vector in the low-temperature monoclinic phase, see table~\ref{tab:character_table}. 
Note that (0.5\,0.5\,0)$_m$ is equivalent to (0.5\,-0.5\,0)$_m$ in the primitive cell and that even the
[110]$_m$ and [1$\bar{1}$0]$_m$ directions are equivalent, because the monoclinic axis is $b_m$. Since the $c_m$ lattice constant is elongated with respect to the average of all three directions, the observation of the magnetic peak (1\,0\,0)$_c$ at larger (absolute) 2$\theta$ values rules out the (0\,0\,1)$_m$ propagation vector at zero field. Assuming a single-$k$ structure, the equivalence of [110]$_m$ and [1$\bar{1}$0]$_m$ directions is broken by the magnetic order which results in a triclinic space group. This is confirmed by the group theory analysis indicating the triclinic space group $P_S\bar{1}$ to be the only maximal magnetic subgroup for this propagation vector~\cite{maxmag}. There is geometric frustration in spite
of the already reduced symmetry in the monoclinic phase.
If a single $ab$ plane (with respect to $P2_1/n$) is ordering antiferromagnetically the neighboring plane at $z$=0.5 remains fully frustrated, and only the breaking of monoclinic symmetry lifts this frustration. 
This is similar to magnetic order associated with a tetragonal to orthorhombic transition in LaOFeAs~\cite{cruz2008,qureshi2010} and to many other materials~\cite{komarek2009,Valldor_BaMn2O3,Niesen2013}. 
The triclinic distortion implied by the magnetic order is visible in the strong thermal expansion anomaly occurring at $T_{\rm N}$, see Fig.~\ref{fig:dlzero}. 
In the magnetic phase with magnetic space group $P_S\bar{1}$, the lattice shrinks along the direction of measurements, as the weak pressure of the apparatus favors again the domain orientations with shorter distances along this direction.
This effect is induced by the ferroelastic character of the structural transition that accompanies the magnetic order in \krc . 

\begin{figure}[t]
\centering
\includegraphics[width=0.95\columnwidth]{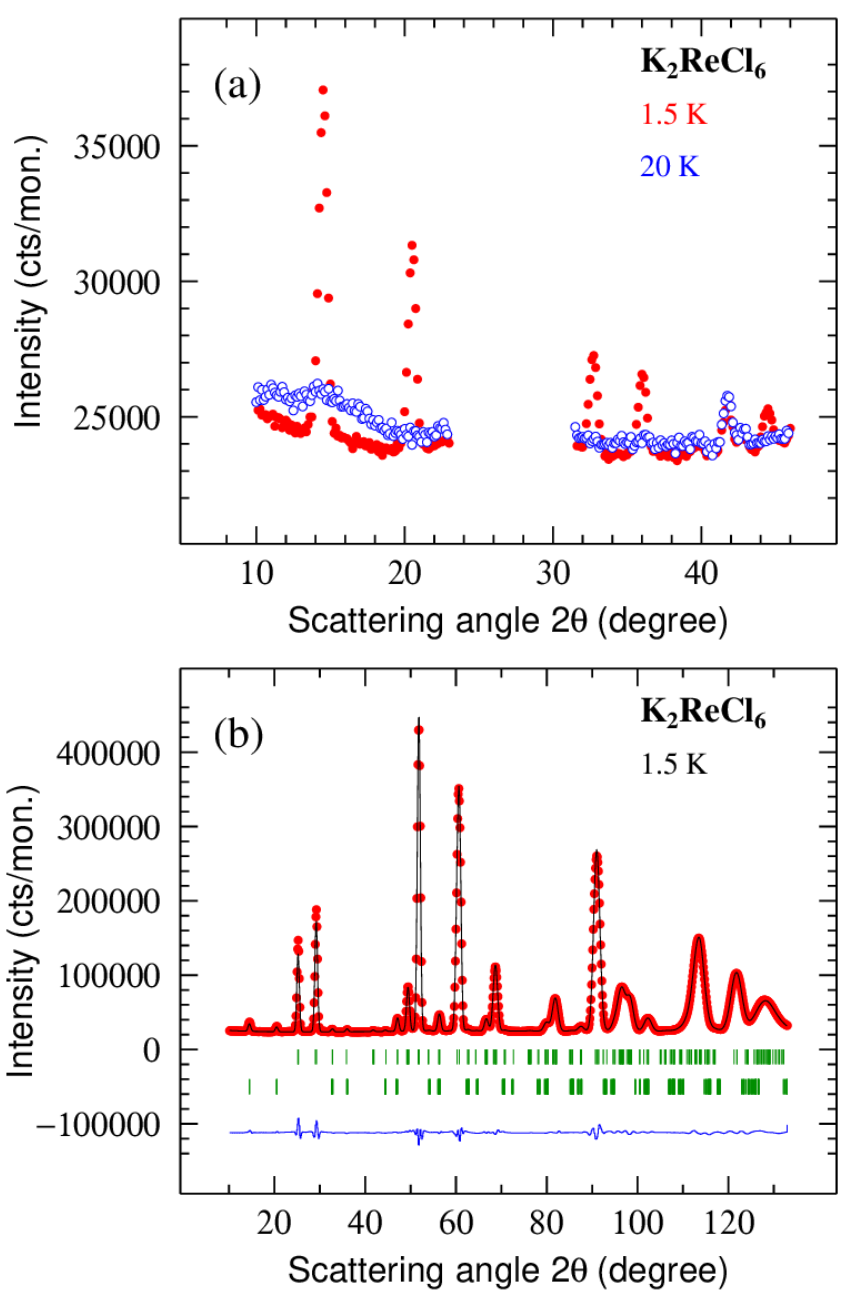}
	\caption{Neutron diffraction powder patterns measured on the DMC diffractometer. (a) Comparison of patterns measured at 20~K in the paramagnetic phase, and at 1.5~K in the ordered phase. Strong nuclear reflections have been excluded for clarity in the 2$\theta$ range 24-31$^{\circ}$. (b) Rietveld refinement of the neutron diffraction pattern measured at 1.5~K. The red circles indicate the experimental data points, the solid black line is a fit including nuclear and magnetic phases modeled with the parent space group $P2_1/n$, and their corresponding Bragg peak positions are indicated by the upper and lower row of green bars, respectively. The blue curve illustrates the difference between fit and data set.}
	\label{fig:DMC}
\end{figure}

In order to determine the zero-field magnetic structure we performed a neutron powder-diffraction experiment, because the occurrence of domains would considerably reduce the precision of a single-crystal experiment. The emergence of magnetic Bragg peaks are shown in Fig.~\ref{fig:DMC}(a) in a small scattering-angle range with the comparison of patterns measured at 20K in the paramagnetic phase and at 1.5K in the ordered state. In the latter pattern, five purely magnetic Bragg peaks are clearly visible and could be indexed with the magnetic propagation vector $k_{\rm mag}=$(0.5\,-0.5\,0)$_m$, and in the former pattern taken slightly above the N\'eel temperature, critical magnetic fluctuations are visible, e.g., around the first magnetic Bragg peak position.


A Rietveld refinement with the 1.5~K powder-diffraction data set is displayed in Fig~\ref{fig:DMC}(b). We describe the data with the nuclear structure in the monoclinic space group $P2_1/n$ (unique axis $b_m$) and a magnetic phase with two magnetic Re ions being purely antiferromagnetically coupled.
The Rietveld analysis also excludes (0\,0\,1)$_m$ as propagation vector of the zero-field magnetic structure corroborating the single-crystal analysis. As discussed above, the (0.5\,-0.5\,0)$_m$ propagation vector causes a further symmetry reduction to a triclinic space group, in which the two Re ions correspond to different orbits. However, the magnetic moments on these two sites could not be independently refined.
The triclinic distortion of the structure is very small, so that splitting of the magnetic orbit is irrelevant.
While the Rietveld analysis can precisely determine the size of the ordered moment, refinements with different orientations of the ordered moments differ only slightly.
The best agreement is obtained with the ordered magnetic moment $m=1.98(8)\mu_{\rm B}$ in the $a_m$-$b_m$ plane, with a canting angle from $b_m$ $\theta=-24(7)^{\circ}$ (or equivalently the magnetic moment projections along the monoclinic axis $m_{a_m}$=0.8(2)$\mu_{\rm B}$, $m_{b_m}$=-1.8(2)$\mu_{\rm B}$ and $m_{c_m}$=0$\mu_{\rm B}$). Refinements modeling the magnetic moment either in the $a_m$-$c_m$ or $b_m$-$c_m$ planes yield $R_{wp}$-factors ($R$ weighted points) slightly larger for both models (4.57\% against 4.56\%) ruling out these solutions. Note that the impact of the magnetic model on $R_{wp}$ is small, as we refine the entire pattern, which is mostly determined by the nuclear peaks. 
Refining all three magnetic components yields an $m_{c_m}$ value much smaller than its error so that this 
coefficient was not varied in the final refinements.
The magnetic structure at zero field incorporated to the low-temperature monoclinic lattice $P2_1/n$ is illustrated in Fig.~\ref{fig:struc}(c), and the temperature dependence of the magnetic moment squared is included in Fig.~\ref{magn-peaks}(a). 
The temperature dependence of $m^2$ and that of the magnetic Bragg-peak intensities measured on the single crystal perfectly agree.

Symmetry analysis can be also applied to the high-field weakly ferromagnetic phase that is stable only in a small pocket of the $B$--$T$ phase diagram. 
Since the applied magnetic fields introduce only a moderate energy scale, one may assume that the
propagation vector in the fundamental cubic lattice remains identical, i.e.\ (0\,0\,1)$_{c}$. 
As (0.5\,0.5\,0)$_{m}$ does not exhibit a weakly ferromagnetic component we obtain thus (0\,0\,1)$_m$ as the propagation vector
in monoclinic units.
The symmetry analysis is more simple for the (0\,0\,1)$_m$ case, which corresponds to the zone center $\Gamma$ in the primitive $P2_1/n$ lattice.
According to representation analysis~\cite{Carvajal1993} or to the maximum magnetic subgroup analysis~\cite{maxmag}, there are two possible magnetic structures describing the magnetic moments at the two sites (0,0,0) and (0.5,0.5,0.5) in $P2_1/n$. 
In magnetic space group 
$P2_1'/c'$ moments are $(m_x,m_y,m_z)$ and $(m_x,-m_y,m_z)$, respectively. In $P2_1/c$, moments are $(m_x,m_y,m_z)$ and $(-m_x,m_y,-m_z)$, respectively.
This can be easily understood because the glide-mirror plane transforms the $m_x$,$m_z$ and $m_y$ components differently. It is obvious that there is a weak
ferromagnetic moment in both cases, which is inconsistent with the drop of magnetic susceptibility at low fields applied along the cubic [1\,1\,1]$_c$ direction, see Fig.~\ref{fig:chi}. This further corroborates our conclusion that the low-field magnetic structure exhibits the (0.5\,0.5\,0)$_m$ propagation vector.
However, this weak ferromagnetism can perfectly explain the anomalous magnetoelastic effects appearing at larger fields, as discussed in section~\ref{MS}.
Because this phase is accompanied with a strong elongation parallel to the field, the magnetic structure of this phase must correspond
to $P2_1'/c'$ which exhibits a weak ferromagnetic moment parallel to the monoclinic $c_m$ direction. This magnetic structure is illustrated in Fig.~\ref{fig:struc}(a).

When the weakly ferromagnetic phase with alignment of $c_m$ directions along the field is left towards higher temperatures, the elongation is only
slowly reverted and full conversion requires heating above the 77\,K phase transition, as it is discussed above. However, when the
temperature is lowered, the abrupt appearance of the triclinic distortion associated with the low-field magnetic structure favors a step-wise rearrangement. In addition, the
magnetic field in the ordered state favors domains with the staggered moments pointing perpendicular to it.

\section{Conclusions}

\krc\ exhibits a series of structural phase transitions associated with rotations of the ReCl$_6$ octahedra, similar to many other members of the antifluorite family. 
In accordance with the symmetry reduction from cubic to monoclinic, single crystals exhibit a ferroelastic multidomain structure with 12 different domain orientations in the monoclinic phase. On the one hand, an efficient partial detwinning can be reached by applying uniaxial pressure along a cubic [1\,0\,0]$_c$ directions because the different structural domains align with their shortest directions along the pressure direction. 
On the other hand, the application of moderate magnetic fields at low temperature yields a very strong elongation of up to 0.6~\%, which results from
the rearrangement of structural domains induced by the external field. The origin of this drastic magnetostructural effect lies in a magnetic-field induced change of the antiferromagnetic structure of \krc .  The zero-field magnetic order is purely antiferromagnetic with vanishing net magnetization, but at high fields a distinct phase with a finite weak ferromagnetic component emerges in a pocket of the $B$--$T$ phase diagram. 
When uniaxial pressure and magnetic field are applied along the same cubic [1\,0\,0]$_c$ direction, their impact on the domain orientation competes with each other and results in a complex interplay of domain switching and memory effects. 

The data used in this article are available from Zenodo~\cite{bertin_2024_10659116}. 

\begin{acknowledgments}
We acknowledge support by the DFG (German Research Foundation) via Project No. 277146847-CRC 1238 (Subprojects A02, B01, and B04)
and by the Bundesministerium f\"ur Bildung und Forschung, Project No. 05K19PK1. This work is partially based on experiments performed at the Swiss spallation neutron source SINQ, Paul Scherrer Institute, Villigen, Switzerland.
\end{acknowledgments}

\end{document}